\documentclass[prb, aps, amsfonts, amsymb, amsmath]{revtex4}

\usepackage{graphicx}

\def\beq{\begin{equation}}
\def\eeq{\end{equation}}
\def\be{\begin{equation}}
\def\ee{\end{equation}}
\def\hm{\hspace{1em}}

\def\iomn{i\omega_n}

\def\t{\mbox{tr}\,}
\def\cG0{{\cal G}_0}
\def\cG{{\cal G}}

\def\spinup{\uparrow}
\def\spindown{\downarrow}
\def\bra{\langle}
\def\ket{\rangle}

\def\vk{{\bf k}}

\def\vr{{\bf r}}

\def\vR{{\bf R}}

\def\e{\varepsilon}

\def\l{\lambda}

\def\s{\sigma}

\def\hn{\hat{n}}

\def\uc2{$U_{c2}$}
\def\uc1{$U_{c1}$}

\def\dS{\Delta\Sigma\,}

\def\ce2o3{$\rm{Ce}_2\rm{O}_3$}

\begin{document}

\title{Self-consistency over the charge-density in
dynamical mean-field theory: a linear muffin-tin implementation and
some physical implications.}

\author{L. V. Pourovskii$^1$, B. Amadon$^2$, S. Biermann$^1$,  and A. Georges$^1$}
\affiliation{$^1$ Centre de Physique Th{\'e}orique, Ecole Polytechnique, CNRS,
91128 Palaiseau Cedex, France\\
$^2$ D{\'e}partement de Physique Th{\'e}orique et Appliqu{\`e}e, CEA,
B.P.~12, 91680 Bruy{\`e}res-le-Ch{\^a}tel, France}

\date{\today}

\begin{abstract}

We present a simple implementation of the dynamical mean-field theory approach to
the electronic structure of strongly correlated materials. This implementation
achieves full self-consistency over the charge density, taking into account correlation-induced
changes to the total charge density and effective Kohn-Sham Hamiltonian.
A linear muffin-tin orbital basis-set is used, and the charge density is computed from moments of the
many body momentum-distribution matrix.
The calculation of the total energy is also considered, with a  proper treatment of
high-frequency tails of the Green's function and self-energy. The method is
illustrated on two materials with well-localized $4f$ electrons, insulating cerium
sesquioxide \ce2o3 and the $\gamma$-phase of metallic cerium, using the Hubbard-I
approximation to the dynamical mean-field self-energy. The momentum-integrated spectral function and
momentum-resolved dispersion of the Hubbard bands are calculated, as well as
the volume-dependence of the total energy.
We show that full self-consistency over the charge density, taking into account its modification
by strong correlations, can be important
for the computation of both thermodynamical and spectral properties, particularly in the
case of the oxide material.

\end{abstract}

\maketitle

\newpage

\section{Introduction}

While density functional theory (DFT) \cite{koh99,jon89} in conjunction
with the local density approximation (LDA) is remarkably successful
in predicting ground-state properties of a wide
range of real materials, it has been found unable to provide the correct description of so-called strongly correlated materials (transition
metal oxides, many actinide and lanthanide-based materials, high T$_c$ superconductors) even
on a qualitative level. In order to overcome these shortcomings of the traditional DFT-LDA scheme, a combination of the DFT-based band structure
techniques with dynamical mean-field theory (DMFT) \cite{geo96} has been proposed \cite{ani97,lichtenstein_lda+dmft_1998,bier2006}. In DMFT one introduces a strong
local Coulomb interaction acting between electrons of a correlated band (for example, the $d-$band in transition-
metal oxides or the $f-$band in actinides). The self-energy is found by first mapping a
full solid onto a quantum impurity model involving a single atom hybridized with an effective bath,
and solving this effective model using many-body techniques.
This self-energy is then promoted to all lattice sites, therefore restoring the translational invariance of
the crystal. As a result one obtains the fully interacting Green's function for the system, from which a wide range of
properties can be extracted.

Starting from the end of the 90's, this new LDA+DMFT approach
to the electronic structure of strongly correlated materials has been rapidly developing, and a number of
different implementations have
been proposed and applied to calculations of the spectral and, in some cases, thermodynamic properties,
of Mott insulators, ferromagnetic 3-$d$
metals, Ce, Pu, and other actinide systems (for reviews, see \cite{georges_strong,kotliar_review,hel02}).
Up to date most LDA+DMFT calculations have been performed using a partially self-consistent scheme,
where the local self-energy is obtained
from a DMFT calculation with the {\it fixed} LDA charge density, and, hence,
with a {\it fixed} LDA Hamiltonian. Therefore, in this simplified scheme one neglects the
impact of the strong on-site Coulomb interaction on the charge
distribution.

Implementing full self-consistency over the charge density is a somewhat
delicate task. Several fully self-consistent LDA+DMFT schemes have been discussed in the literature 
\cite{sav04,min05,lech06,anis07}, while 
only two actual implementations have appeared up to date. The first one,
due to Savrasov and Kotliar~\cite{sav04},
is based on the full-potential LMTO method. In order to compute the LDA+DMFT charge density
one has to construct from the LDA Hamiltonian and the DMFT local self-energy a local Green's function
of the interacting system.
In Ref.~[\onlinecite{sav04}]
this task has been accomplished by first finding the (right- and left-) eigenfunctions and
eigenvectors of the Kohn-Sham Hamiltonian combined with the self-energy
$H_{KS}(\vk)+\Sigma(i\omega)$ and
then performing the Brillouin zone integration by means of the
tetrahedron method \cite{lam84}. As the self-energy is a complex
function, this scheme requires the
diagonalization of a non-Hermitian matrix for each $\vk-$point and Matsubara frequency,
which is a computationally demanding task. Another
scheme has been proposed by Minar {\it et al.} \cite{min05} on the basis of the KKR
Green's function method. Their approach requires solving the radial Dirac
equation with the self-energy being added to the LDA atomic potential. Generally the self-energy is a
non-diagonal matrix, therefore an additional coupling
is generated between Dirac equations for different magnetic
quantum numbers $m$, making the solution of the Dirac equation a highly
non-trivial task. Minar {\it et al.} applied their scheme to calculations of the
FeNi binary alloy, which forms a cubic lattice. Hence, in this case, both the Green's function
and self-energy are diagonal in the orbital indexes, and there is no
additional coupling generated in the Dirac equation. However, this problem
will certainly arise for lower symmetry lattices or $f-$electron compounds.

In the present article, we propose and apply a relatively simple implementation of
the fully self-consistent LDA+DMFT method. The charge density
(including correlation effects described by the many-body self-energy)
is expressed in terms of a $\vk$-dependent momentum-distribution matrix
$N_{LL'}^\vk$, which is obtained from the many-body Green's function
by summing over frequency. Furthermore, using a linear muffin-tin orbital basis set,
we reduce the calculation of the charge density to that of
three moments, involving this matrix and the Kohn-Sham Hamiltonian itself.
In order to calculate the total energy, a functional
of both the charge density and the on-site components of the Green's function
associated with local orbitals can be used~\cite{sav04,georges_strong},
as previously discussed by Savrasov and Kotliar.
In all these calculations, obtaining the charge
density and total energy with sufficient accuracy requires a careful treatment of
the high-frequency tails of
the Green's function and self-energy, and we derive here appropriate formulas
to handle this issue.

In order to illustrate this approach, and to assess the importance of full self-consistency
on the charge density, we perform in this article fully-selfconsistent LDA+DMFT
calculations of the density of states, the band structure, and the volume-dependence of the
total energy of two materials: $\gamma$-cerium and cerium
 sesquioxide Ce$_2$O$_3$. The quasi-localized $f$-electrons are treated using DMFT in conjunction
 with the Hubbard-I approximation as a quantum impurity solver~\cite{hubbard_1}.
 Our theory reproduces the experimentally observed splitting of the rare-earth $f$-band into
occupied lower and empty upper Hubbard bands, while the conventional LDA calculations incorrectly predict this band to be pinned at the
Fermi level \cite{skr01}. We show that the self-consistency over the charge density
shifts significantly the positions of the Hubbard bands (in comparison to using a
frozen LDA charge density), in both $\gamma$-Ce
and Ce$_2$O$_3$. It is therefore important for the correct description of
spectral and optical properties of those materials.

The paper is organized as follows. In Section~\ref{sec:method},
we describe our self-consistent LDA+DMFT implementation.
After a brief reminder of the basic LDA+DMFT scheme (Sec.~\ref{lda+dmft-scheme})
and of the LMTO basis-set (Sec.~\ref{sec:lmto}), we
derive the expression for the charge density in this basis (Sec.~\ref{sec:density}).
After reviewing the total energy functional of LDA+DMFT (Sec.~\ref{sec:energy_lda+dmft}),
we discuss the practical calculation of the energy (Sec.~\ref{sec:energy_practical}).
We conclude Sec.~\ref{sec:method} by describing the specific form of
the interaction vertex used in this article, as well as the double
counting correction and the Hubbard-I impurity solver (Sec.~\ref{sec:interaction_solver}).
Section~\ref{sec:results} presents the results for Ce$_2$O$_3$ (Sec.~\ref{sec:ce2o3})
and $\gamma$-Ce (Sec.~\ref{sec:cerium}).

\section{Implementation of the fully self-consistent LDA+DMFT method}
\label{sec:method}

\subsection{The LDA+DMFT formalism: a brief reminder}
\label{lda+dmft-scheme}

The LDA+DMFT approach to electronic structure is based on two key quantities:
the charge density $\rho(\vr)$ and the local Green's function of the solid,
or more precisely the projection $G_{ab}(\omega)$ of the full Green's function onto a
single atomic site and on the subspace of correlated orbitals.
Both quantities have to be determined self-consistently, following an iterative cycle
which is summarized on Fig.~\ref{fig:lda+dmft_loop}. This can be rationalized as a
functional of both $\rho(\vr)$ and $G_{ab}$, as detailed later in Sec.~\ref{sec:energy_lda+dmft}.

Let us follow this iterative cycle, starting from a charge density profile $\rho(\vr)$.
A Kohn-Sham (KS) potential is constructed from $\rho(\vr)$ as:
\beq
v_{KS}(\vr) = v_c(\vr) + v_{H}\left[\rho(\vr)\right] + v_{xc}\left[\rho(\vr)\right]
\label{eq:KSpot_1}
\eeq
in which $v_c$ is the periodic potential of the (fixed) ions,
$v_H=\int d\vr' \frac{e^2}{\vert\vr-\vr'\vert}\rho(\vr')$ is the Hartree
potential, and $v_{xc}=\delta E_{xc}/\delta\rho(\vr)$ is the exchange-correlation
potential.
Hence, the functional dependence of $v_{KS}$ on $\rho(\vr)$ is kept identical
to that of conventional DFT. In practice, $v_{xc}$ will be computed from $\rho(\vr)$
using the LDA form of the exchange-correlation energy
$E_{xc}^{LDA}=\int d\vr \rho(\vr)\varepsilon[\rho(\vr)]$, with $\epsilon[\rho]$ the
energy density of the homogeneous electron gas.
Solving the single-particle Schr\"{o}dinger equation associated with $v_{KS}$ yields
the KS eigenenergies $\epsilon_{\vk\nu}^{KS}$ and eigenfunctions $|\vk\nu\rangle$
(with $\nu$ a band index), forming
the KS effective one-particle hamiltonian:
\beq
H_{KS}\,=\,\sum_{\vk\nu}\,\epsilon_{\vk\nu}^{KS}\,\vert\vk\nu\rangle\langle\vk\nu\vert
\label{eq:def_KS}
\eeq

At this stage, it is convenient to introduce a set of localized basis functions
$\chi_{L \vR}(\vr)$, where $\vR$ denotes an atomic position, and $L$
stands for all orbital indices (e.g., $L=\{l,m,\sigma\}$).
In the following, we shall consider linearized muffin-tin orbitals (LMTOs),
but different basis-sets can be used and have been considered
in the literature~\cite{pav04,ani05,lech06} (e.g., Wannier functions).
%
The electron creation operator at a point $\vr$ in the solid can be
expanded on this basis as:
\beq
\psi^\dagger(\vr) = \sum_{\vR,L} \chi^*_{L\vR}(\vr)\,c_{L\vR}^\dagger
\eeq
and the KS hamiltonian reads:
\beq
H_{KS}\,=\,\sum_{\vk L} H^{KS}_{LL'}(\vk) c^\dagger_{\vk L}c_{\vk L'}
\label{eq:KS_lmto}
\eeq

In the LDA+DMFT approach, this Hamiltonian is supplemented by many-body terms.
These many-body terms act in the subspace generated by a set of orbitals
corresponding in practice to the orbitals (e.g
$d$ or $f$-orbitals) for which a description beyond DFT-LDA is needed
(note however that all other orbital components in the valence will also
be modified indirectly by feedback effects of the self-energy associated with
the correlated ones).
For simplicity, we restrict the discussion here to the case
of one `correlated' atom per unit cell.
The orbitals generating the correlated subset need not coincide with basis
functions in general~\cite{lech06}. However in the present work, we do choose them as
a specific subset $\chi_{a\vR}(\vr)$ of the LMTOs. Hence, $L=\{l,m,\sigma\}$ runs over
all orbitals retained in the valence, while $a=\{m,\sigma\}$ runs only over
the `correlated' subset (denoted by $\cal{C}$ in the following).
The many-body Hamiltonian considered in LDA+DMFT reads:
\beq
H\,=\,H_{KS}-H_{DC}+H_U
\label{eq:mb_ham}
\eeq
In this expression, $H_{DC}$ (corresponding to a one-body potential $V_{DC}$)
is a double-counting correction. Indeed, some of the
local Coulomb interaction effects are already taken into account in the
exchange-correlation energy, and hence in $H_{KS}$.
The many-body terms $H_U$ act in the subset of correlated orbitals
only. They correspond to matrix elements of the Coulomb interaction, and will in general
involve general 2-particle terms of the form
$\sum_{\vR,abcd} U_{abcd} c^\dagger_{a\vR}c^\dagger_{b\vR}c_{d\vR}c_{c\vR}$.
(In the present work however, only density-density terms will be retained:
the form of $H_U$ and $H_{DC}$ used in this article is discussed further
in Sec.~\ref{sec:interaction_solver}).

Let us consider the full Green's function of the solid
$G(\vr,\vr';\tau-\tau')\equiv - \bra T\psi(\vr,\tau)\psi^\dagger(\vr',\tau')\ket$,
which can be decomposed on the basis set as:
\beq
G(\vr,\vr';\tau-\tau') = \sum_{\vR\vR'}\sum_{LL'} \chi_{L\vR}(\vr)\,
G_{LL'}(\vR-\vR',\tau-\tau')\,\chi_{L'\vR'}(\vr')^*
\label{eq:green_basis}
\eeq
DMFT focuses on the local components $G_{ab}$ of the Green's function,
on the same atomic site ($\vR=\vR'$) and within the correlated subspace.
The key idea (which can be viewed as a representability assumption) is
that $G_{ab}$ can be represented by an effective
local model, which is a multi-band
generalization of an Anderson impurity model described by the effective action:
\begin{widetext}
\begin{equation}
S=
- \int^{\beta}_{0} d\tau \int^{\beta}_{0} d\tau'\, \sum_{ab}
c^{\dagger}_{a}(\tau)\,[{\cal G}_0^{-1}]_{ab}(\tau-\tau')\,c_{b}(\tau')
+
\int d\tau\, H_U
\label{Smulti}
\end{equation}
\end{widetext}
In this expression, ${\cal G}_0$ is the dynamical mean-field, analogous to the
familiar Weiss mean-field in classical mean-field theory, the key difference being
that here it is a frequency-dependent (i.e energy-scale dependent) quantity. It can
also be viewed as the hybridization function which connects the correlated atom
(effective impurity) to its environment:
$\Delta_{ab}(z)=z\,\delta_{ab}-\epsilon^f_{ab}-[{\cal G}_0(z)^{-1}]_{ab}$. In this expression,
$z$ is an arbitrary frequency in the complex plane, and $\epsilon_f$ is a matrix
of effective on-site atomic levels (see sec.~\ref{sec:interaction_solver}).
The dynamical mean-field ${\cal G}_0$ (or $\Delta$) is determined from a self-consistency
condition, which expresses that the impurity-model Green's function faithfully represents the
local Green's function in the solid projected onto the correlated subset, and hence that the
two quantities should coincide. Furthermore, an approximation is made, namely that the
many-body self-energy has components on the basis set which are (i) local and (ii) non-zero only in
the correlated subspace, so that it also coincides with its impurity model counterpart and
takes the form:
\beq
\Sigma^{\vR\vR'}_{LL'}(z)\,=\,\delta_{\vR,\vR'}\,
\left(%
\begin{array}{cc}
  0 & 0 \\
  0 & \Sigma^{imp}_{ab}(z) \\
\end{array}%
\right)
\label{eq:sigma_matrix}
\eeq
The impurity model Green's function and self-energy are defined as:
\beq
G_{ab}^{imp}(\tau-\tau') \equiv - \bra Tc^\dagger_a(\tau)c_b(\tau')\ket_{imp}\,\,\,,\,\,\,
\Sigma^{imp}_{ab} \equiv [{\cal G}_0^{-1}]_{ab} - [G_{imp}^{-1}]_{ab}
\label{eq:imp_G_Sigma}
\eeq
in which the average indicated by $\bra ... \ket_{imp}$ is taken with respect
to the effective action (\ref{Smulti}). The self-consistency condition which determines
${\cal G}_0$ can thus be expressed in a concise way as:
\beq
\hat{P}_{\vR}^{\cal{C}}\, \hat{G} \, \hat{P}_{\vR}^{\cal{C}} \,=\,\hat{G}_{imp}
\label{eq:scc_concise}
\eeq
in which the full Green's function $\hat{G}$ of the solid is projected on a given
correlated atom, and onto the correlated subset with the projector:
\beq
\label{project}
\hat{P}_{\vR}^{\cal{C}}\equiv\sum_{a}|\chi_{\vR a} \rangle \langle \chi_{\vR a}|
\eeq
Given (\ref{eq:imp_G_Sigma}) and given Dyson's equation relating $\hat{G}$ and
$\Sigma$, Eq.~(\ref{eq:scc_concise}) yields an implicit relation between $G_{imp}$
and ${\cal G}_0$. To be fully explicit, let us go through the iterative process
(the DMFT loop depicted in Fig.~\ref{fig:lda+dmft_loop}) which determine
$G_{imp},\Sigma_{imp}$ and ${\cal G}_0$ self-consistently. Given an initial guess
for ${\cal G}_0$, the impurity Green's function $G_{imp}$  is calculated using some appropriate
`impurity solver', and the impurity self-energy is obtained as:
$\Sigma^{imp}_{ab} \equiv [{\cal G}_0^{-1}]_{ab} - [G_{imp}^{-1}]_{ab}$, and used as the
only non-zero block of $\Sigma_{LL'}$ (Eq.~\ref{eq:sigma_matrix}). The Green's function
in the solid can then be calculated as:
\beq
[G^{-1}]_{LL'}(\vk,z) = (z+\mu)\delta_{LL'} - H^{KS}_{LL'}
+ V^{DC}_{LL'} - \Sigma_{LL'}(z)
\label{eq:gll'}
\eeq
and summed over $\vk$-points in the Brillouin zone
in order to yield the local Green's function on the atomic site associated with
the correlated atom as:
\beq
G_{loc}(z)\,=\,\sum_{\vk}\,
\left[(z+\mu)\delta_{LL'} - H^{KS}_{LL'}
+ V^{DC}_{LL'} - \Sigma_{LL'}(z)\right]^{-1}
\label{eq:scc_esc_dmft}
\eeq
The block corresponding to the correlated orbitals is extracted from this expression
and inverted in order to get an iterated value for the dynamical mean-field, as:
\beq
[{\cal G}_0^{-1}]_{ab}=\Sigma_{imp}+[P^{\cal{C}}G_{loc}P^{\cal{C}}]^{-1}\,,
\label{eq:new_G0}
\eeq
and the DMFT iteration can be pursued.
Note that, even though the self-energy matrix has only components in the
subspace of correlated orbitals, components of the Green's function corresponding to all
valence orbitals ($s,p,\cdots$) are modified due to the matrix
inversion. Hence there is a feedback of correlation effects onto these orbitals as well.

At the end of the DMFT cycle, the resulting Green's function can be considered as a full
many-body solution of the Hamiltonian defined by Eq.~(\ref{eq:mb_ham}), for
a given charge density determining $H_{KS}$,
under the approximation that the corresponding self-energy is taken
to be purely local.
The charge density can then be recalculated from the Green's function at the end
of this cycle, as:
\beq
\rho(\vr)\,=\,\sum_{\vR\vR'}\sum_{LL'} \chi_{L\vR}(\vr)\,
G_{LL'}(\vR-\vR',\tau-\tau'=0^-)\,\chi_{L'\vR'}(\vr)^*
\label{eq:charge_from_Green}
\eeq
From this updated charge density, a new Kohn-Sham potential is constructed from
(\ref{eq:KSpot_1}), and the associated one-particle Schr\"{o}dinger equation is
solved again in order to produce an updated $H_{KS}$, which serves as a new input
for the DMFT cycle. This is indicated as the DFT loop on Fig.~\ref{fig:lda+dmft_loop}.
Full self-consistency is reached when all local quantities
($G_{imp}$,$\Sigma_{imp}$,${\cal G}_0$) as well as the charge density $\rho(\vr)$, have converged.
Note that the LDA+DMFT self-consistent charge density is affected
by correlation effects through the many-body self-energy and that it differs from its
LDA self-consistent value $\rho_{LDA}(\vr)$. Furthermore, it also differs from the
charge density associated with the occupied Kohn-Sham orbitals and evaluated with
the converged $H_{KS}$ but $\Sigma=0$ in (\ref{eq:charge_from_Green}), i.e:
$\rho_{KS}(\vr)\equiv\sum_{\vk\nu}^\prime |\psi_{\vk\nu}(\vr)|^2$. That is, the Kohn-Sham
representation is not used for the self-consistent charge density within
LDA+DMFT.
In the next two sections, we first give a brief reminder of the LMTO formalism
and basis-set, and then proceed with the practical evaluation of expression
(\ref{eq:charge_from_Green}) for the charge density in this basis-set.

\begin{widetext}
\begin{center}
\begin{figure}[t]
\includegraphics[width=10 cm]{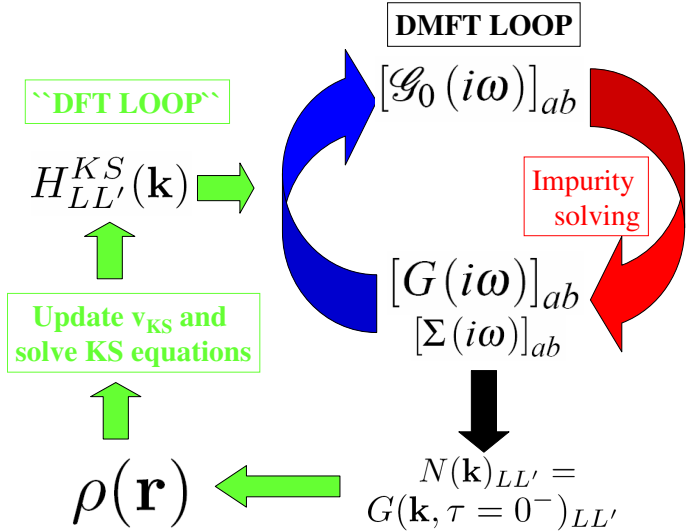}
\caption{(Color online). DMFT combined with electronic structure calculations. Starting from a local electronic
density $\rho(\vr)$, the associated Kohn-Sham potential is calculated and the Kohn-Sham equations
are solved. The Kohn-Sham hamiltonian $H^{KS}_{LL'}(\vk)$ is expressed in a localised basis set
(e.g LMTOs).
A double-counting term is subtracted to obtain the effective
one-electron Hamiltonian $H_0\equiv H^{KS}-H^{DC}$. The local self-energy matrix for the subset of
correlated orbitals is obtained through the iteration of the DMFT loop: a multi-orbital impurity
model for the correlated subset is solved (red (lighter grey) arrow), containing as an input the dynamical mean-field
(or Weiss field ${\cal G}_0$). The self-energy $\Sigma_{ab}$ is combined with $H_0$ into
the self-consistency condition Eqs.~(\ref{eq:scc_esc_dmft},\ref{eq:new_G0})
in order to update the Weiss field (blue (darker grey) arrow).
At the end of the DMFT loop, the components of the full, $\vk$-dependent,
Green's function in the local basis set can be calculated, yielding the momentum-distribution
matrix $N(\vk)_{LL'}$ and the updated charge
density $\rho(\vr)$ as described in Sec.~\ref{sec:density}. This updated
charge density is used to compute the new Kohn-Sham potential until
a converged local density is also reached (for the DFT loop). At the same time,
the chemical potential must also be adjusted self-consistently with the
total number of electrons.
}
\label{fig:lda+dmft_loop}
\end{figure}
\end{center}
\end{widetext}

\subsection{The LMTO basis set}
\label{sec:lmto}

As discussed in detail in section \ref{lda+dmft-scheme},
since DMFT emphasizes local correlations, one needs to construct
a localized basis set, i.e basis
functions which are centered on the atomic positions $\vR$ in the crystal lattice.
Up to now, most implementations have used basis sets based on
linear muffin-tin orbitals~\cite{andersen_lmto_1975_prb} (LMTOs)
$\chi_{L\vR}(\vr)=\chi_L(\vr-\vR)$.
These basis sets offer the advantage to carry over the physical intuition
of atomic orbitals from the isolated atom to the solid.
The LMTO method has been extensively used in electronic structure calculations and
thoroughly described in review articles~\cite{skriver_book,andersen_nmto_2000_wshop}.
Here we only outline the
main features of the LMTO basis set which
are relevant to the selfconsistent implementation of the LDA+DMFT scheme.

MT-methods suppose that the crystal potential is spherically symmetric near each atomic site and constant in the
interstitial region between atoms.  The conventional LMTO method employs a further simplification,
the atomic-sphere approximation (ASA), which assumes
that the whole space of a crystal is filled with atomic spheres and neglects both the overlap and interstitial spaces.
The LMTO basis is constructed from the solution (partial wave) $\phi_{L\vR}$ of the spherically symmetric
KS potential inside the atomic sphere located at $\vR$ for a certain energy $E_{\nu}$
(typically, the center of gravity of a band)
and its energy derivative $\dot{\phi}_{L\vR}$.
The angular dependence is provided by a spherical
harmonic $Y_L$, corresponding to the
orbital and magnetic quantum numbers L).
The expression for a linearized MT-orbital is:
\begin{equation}\label{hi_lmto}
\chi_{L\vR}^{\alpha}(\vr)=\phi_{L\vR}(\vr)+\sum_{L'\vR'}\dot{\phi}_{L'\vR'}(\vr)h^{\alpha}_{L'\vR'L\vR},
\end{equation}
where the first and second terms in the right-hand side of this equation are usually called the "head" and "tail",
respectively.
The superscript $\alpha$ designates a particular LMTO
representation (the so-called ``screened'' non-orthogonal representation),
which is defined by the choice of the envelope functions for the interstitial.
We have actually employed here the so-called nearly-orthogonal
$\gamma$-representation~\cite{lmto_ref_86},
where $h^\gamma$ is chosen as:
\beq
h^{\gamma}_{L'\vR'L\vR}\,\equiv\,H^{KS}_{L'\vR'L\vR}-E_\nu\,\delta_{\vR'\vR}\delta_{L'L}
\label{def_small_h}
\eeq
Using the orthogonality property of the partial wave and its energy derivative $\langle
\dot{\phi}_{L\vR}|\phi_{L\vR} \rangle =0$ one may easily show \cite{lmto_ref_86} that the head of the LMTO
in the $\gamma$-representation is orthogonal to any LMTO
centered on
any other site up to a second-order contribution
due to the overlap of the energy derivatives of the partial waves.
Then the basis can be made completely orthogonal by means
of an numerical orthogonalization, for example, the L{\"o}wdin transformation \cite{low50}.
The use of an orthogonal basis is technically simpler when implementing LDA+DMFT
(although a non-orthogonal basis may be more advisable in principle since it
is expected to reduce the range of interaction terms).
A numerical orthogonalization usually introduces undesirable
mixing between strongly and weakly correlated states. Because the LMTO basis
in the $\gamma$ representation is already nearly-orthogonal, the inter-orbital mixing are
small however.

One of the main advantages of the LMTO technique is the small size of its basis.
It can be made even smaller through use of the
downfolding procedure \cite{lamb86}, which allows to reduce the size of the Hamiltonian
by folding down those states
which are located well above the valence band. This technique
substantially reduces the computational effort, a gain which is
especially important in the case of the time-consuming
LDA+DMFT calculations. The down-folded LMTO can be expressed
as follows:
\begin{equation}\label{hi_down}
\chi_{L\vR}^{\gamma}(\vr)=\phi_{L\vR}(\vr)+\sum_{L'\vR'}\dot{\phi}_{L'\vR'}(\vr)h^{\gamma}_{L'\vR'L\vR}+\sum_{H\vR''}\phi_{H\vR''}(\vr)Z_{H\vR''L\vR},
\end{equation}
where $L$ labels "active" orbitals, which are explicitly included in the Hamiltonian,
$H$ runs over the set of downfolded orbitals, the matrix $Z$ can be expressed through the structure constants and
potential parameters \cite{lamb86}. Following Ref.~[\onlinecite{lamb86}] we
neglect the energy dependence of the downfolded orbitals
$\phi_{H\vR''}(\vr)$.
In the case of a periodic crystal,
one may rewrite expression (\ref{hi_down}) at each $\vk$-point in the Brillouin zone as:
\begin{equation}\label{hi_down_k}
\chi_{L}^{\vk}(\vr)=\phi_{L}(\vr)+\sum_{L'}\dot{\phi}_{L'}(\vr)h^{\vk}_{L'L}+\sum_{H}\phi_{H}(\vr)Z^{\vk}_{HL},
\end{equation}

\subsection{The Calculation of the Charge Density}
\label{sec:density}

%
In this section, we describe the practical implementation of the
calculation of the LDA+DMFT charge density in the LMTO basis set. Expression
(\ref{eq:charge_from_Green}) can be rewritten as:
\beq
\label{ch_dn}
\rho(\vr)\,=\sum_{LL'} \sum_\vk\,\chi_{L\vk}(\vr)\,
N^{\vk}_{LL'} \chi_{L'\vk}^*(\vr)
\eeq
in which $N^{\vk}_{LL'}$ is the $\vk$-dependent occupancy matrix related to the
full Green's function (\ref{eq:green_basis}) by:
\beq\label{occ_mat}
N^{\vk}_{LL'}\equiv G_{LL'}(\vk,\tau=0^-)=T\sum_{n} G_{LL'}(\vk,i\omega_n)\,e^{i\omega 0^+}
\eeq
The last equality is expressed as a sum over fermionic Matsubara frequencies
$\omega_n=(2n+1)\pi/\beta$ corresponding to the temperature $T=1/\beta$.
By inserting the linear MT-orbitals (\ref{hi_down_k}) in (\ref{ch_dn}), one can obtain
a general formula for the LDA+DMFT charge density in the LMTO framework, which involves
simply the calculation of momentum averages of $N$ and of products of
$N$ with the hamiltonian matrix $h$. This expression can be further simplified
if the atomic-sphere approximation (ASA) is used,
in which the potential is approximated as spherically symmetric within the MT-spheres.
In accordance with this approximation, crossed terms between different angular momentum channels
can be neglected. Neglecting also the overlap between spheres and using the orthogonality
$\langle \phi_{L}|\phi_{L'} \rangle =\delta_{LL'}$ and
$\langle \phi_{L}|\phi_{H} \rangle =0$ between partial waves yields
our final expression for the angular-averaged charge density in a given unit cell
(valid within the ASA):
\beq\label{den_mom}
\rho(r)=\sum_{L} \big ( m_{L}^{(0)}  |\phi_{L}(r)|^2+2
m_{L}^{(1)} \phi_{L}(r)\dot{\phi}_{L}(r)  +  m_L^{(2)}
 |\dot{\phi}_{L}(r)|^2  \big ) + \sum_H m_H^{(0)} |\phi_{H}(r)|^2,
\eeq
In this expression, the moments $m^{(i)}$'s are defined from the $\vk$-dependent
occupancy matrix and LMTO first-order Hamiltonian $h$ (for a given atom $\vR$) as:
\begin{eqnarray}\label{moments}
\nonumber
&&m_L^{(0)}\equiv\sum_{\vk}N^{\vk}_{LL}=\bra\rm{tr} N\ket_\vk\\
&&m_H^{(0)}\equiv\sum_{\vk}\sum_{LL'}Z^{\vk}_{HL}N^{\vk}_{LL'}Z^{\vk}_{L'H}=\bra\rm{tr}(Z_H N Z_H)\ket_\vk\\
\nonumber
&&m_L^{(1)}\equiv\sum_{\vk}\sum_{L'}N^{\vk}_{LL'}h^{\vk}_{L'L}=\bra\rm{tr} (N h)\ket_\vk\\
\nonumber
&&m_L^{(2)}\equiv\sum_{\vk}\sum_{L'L''}h^{\vk}_{LL'}N^{\vk}_{L'L''}h^{\vk}_{L''L}
=\bra\rm{tr} (h N h)\ket_\vk
\end{eqnarray}
Expressions (\ref{den_mom},\ref{moments}) are similar to those used in the context of usual DFT
in the LMTO-ASA formalism, the key difference being that in the DMFT context, the momentum-distribution
matrix $N^\vk$ is computed from the many-body Green's function according to
(\ref{ch_dn}) instead than from filling independent orbitals as in the KS representation of the
density. In practice, the moments (\ref{moments}) are computed
at the end of the DMFT cycle and then passed on into the LMTO
electronic structure part of the program, where a new total charge
density is computed according to formula (\ref{den_mom}),
as indicated on Fig.~\ref{fig:lda+dmft_loop}.


\subsection{The Total Energy Functional}
\label{sec:energy_lda+dmft}

In order to discuss total energy calculations in the LDA$+$DMFT framework,
it is best to use a formulation of this scheme in terms of
a (free-) energy functional.
Kotliar and Savrasov~\cite{kotliar_savrasov_newton,sav04,kotliar_review_rmp_2006}
have introduced for this
purpose a (``spectral-density-'') functional of both the total charge density $\rho(\vr)$
and the on-site Green's function in the correlated subset:
$G_{ab}^{\vR\vR}$ (denoted $G_{ab}$ for simplicity in the following).
Let us emphasize that these quantities are independent, since $G_{ab}$ is
restricted to local components and to
a subset of orbitals so that $\rho(\vr)$ cannot be reconstructed from it.
The functional is
constructed by introducing source terms
$\l(\vr)=v_{KS}(\vr)-v_c(\vr)$
and $\Delta\Sigma_{ab}(\iomn)$
coupling to the operators $\psi^\dagger(\vr)\psi(\vr)$ and to
$\sum_{\vR}\chi^*_a(\vr-\vR)\psi(\vr,\tau)\psi^\dagger(\vr',\tau')\chi_b(\vr'-\vR)
=c_{a\vR}(\tau)c^\dagger_{b\vR}(\tau')$, respectively.
Furthermore, the Luttinger-Ward \cite{luttinger_ward_2_1960} part of the functional is approximated by
that of the on-site
local many-body Hamiltonian $H_U-H_{DC}$ introduced above.
This yields:
\begin{widetext}
\begin{eqnarray}\nonumber
&\Omega[\rho(\vr),G_{ab};v_{KS}(\vr),\dS_{ab}]_{LDA+DMFT}
\,=\\ \nonumber
&-\t\ln[\iomn+\mu+\frac{1}{2}\nabla^2-v_{KS}(\vr)-\chi^*.\dS.\chi]
-\int d\vr\,(v_{KS}-v_c)\rho(\vr) -\t [G.\dS] + \\ \nonumber
&+\frac{1}{2}\int d\vr\,d\vr' \rho(\vr)\,\frac{e^2}{\vert\vr-\vr'\vert}\,\rho(\vr')
+ E_{xc}[\rho(\vr)]
+\sum_\vR\left(\Phi_{imp}[G^{\vR\vR}_{ab}]-\Phi_{DC}[G^{\vR\vR}_{ab}]\right)
\label{eq:LDA+DMFT_func}
\end{eqnarray}
\end{widetext}
In this expression, $\chi^*.\dS.\chi$ denotes the ``upfolding'' of the local
quantity $\dS$ to the whole solid:
$\chi^*.\dS.\chi=\sum_\vR\sum_{ab}\chi^*_a(\vr-\vR)\Sigma_{ab}(\iomn)\chi_b(\vr'-\vR)$.
Variations of this functional with respect to the sources
$\delta\Omega/\delta\,v_{KS}=0$ and $\delta\Omega/\delta\Sigma_{ab}=0$ yield the
standard expression of the local density and local Green's function in terms of the
full Green's function of the solid, which we have used in the previous section:
\beq
\rho(\vr) = \bra\vr|\hat{G}|\vr\ket\,\,\,,\,\,\,
G_{ab}(\iomn) =
\bra\chi_{a\vR}|\hat{G}|\chi_{b\vR}\ket
\label{eq:rho_and_G}
\eeq
with:
\beq
\hat{G} = \left[\iomn+\mu+\frac{1}{2}\nabla^2-v_{KS}(\vr)-\chi^*.\dS.\chi\right]^{-1}
\eeq
Note that these expressions, as well as the functional,
are written here in a manner which does not refer explicitly to a specific basis set.
The formalism does depend, however, on the choice of localized orbitals defining the
correlated subspace.

From these relations, the Legendre multiplier
functions $v_{KS}$ and $\dS$ can be eliminated in terms of
$\rho$ and $G_{ab}$, so that a functional
of these local observables only is obtained:
\begin{widetext}
\beq
\Gamma_{LDA+DMFT}[\rho,G_{ab}]=
\Omega_{LDA+DMFT}\left[\rho(\vr),G_{ab};\l[\rho,G],\dS[\rho,G]\right]
\label{eq:Gamma}
\eeq
\end{widetext}
Extremalisation of this functional with respect to $\rho$
($\delta\Gamma/\delta\rho=0$)
and $G_{ab}$ ($\delta\Gamma/\delta\,G_{ab}=0$) yields the expression of the
Kohn-Sham potential and self-energy correction at self-consistency:
\beq
v_{KS}(\vr) = v_c(\vr) + \int d\vr' \frac{e^2}{\vert\vr-\vr'\vert}\,\rho(\vr') +
\frac{\delta E_{xc}}{\delta\rho(\vr)}
\eeq
\beq
\dS_{ab} = \frac{\delta\Phi_{imp}}{\delta\,G_{ab}}-\frac{\delta\Phi_{DC}}{\delta\,G_{ab}}
\equiv \Sigma^{imp}_{ab}-V^{DC}_{ab}
\eeq
Hence, one recovers from this functional the defining equations of the LDA$+$DMFT
combined scheme detailed in the previous section, including self-consistency
over the local density (\ref{eq:rho_and_G}).
%
%

The free-energy (\ref{eq:LDA+DMFT_func},\ref{eq:Gamma}) leads to the following expression
of the total energy:
\begin{widetext}
\begin{eqnarray}\label{edmft}
E_{LDA+DMFT} & =\sum_{\vk,LL'} H^{KS}_{LL'}(\vk) N^{\vk}_{LL'}+\int d\vr [v_c(\vr)-v_{KS}(\vr)]\rho(\vr)+ \\
\nonumber &+\frac{1}{2}\int d\vr d\vr' \rho(\vr)\, \frac{e^2}{\vert\vr-\vr'\vert}\,
\rho(\vr')+ E_{xc}[\rho]+ \bra H_U \ket - E_{DC}.
\end{eqnarray}
\end{widetext}
The sum of the first four terms in this expression is reminiscent of the expression
of the total energy in DFT, $E_{DFT}[\rho(\vr)]$, evaluated at the self-consistent charge density
$\rho(\vr)$, except for the fact that the many-body (LDA+DMFT)
momentum-distribution matrix enters the first term. Instead, within DFT, the first
term reads $\sum'_{\vk\nu} \e^{KS}_{\vk\nu}$, where the prime indicates that the sum is to
be taken only over occupied KS orbitals (filled up to $\mu_{KS}$, adjusted so that the
total number of electrons is obtained. Hence, in practice, the total energy can be
calculated as:
\begin{eqnarray}\nonumber
E_{LDA+DMFT}\,=\,&&\left(E_{DFT}[\rho(\vr)] - \sum^\prime_{\vk\nu} \e^{KS}_{\vk\nu}\right)+
\sum_{\vk,LL'} H^{KS}_{LL'}(\vk) N^{\vk}_{LL'} + \bra H_U \ket - E_{DC}=\\
 && = E_{c}[\rho]+E_H[\rho]+E_{xc}[\rho] + \sum_{\vk,LL'} H^{KS}_{LL'}(\vk) N^{\vk}_{LL'} + \bra H_U \ket - E_{DC}
\label{eq_energy2}
\end{eqnarray}
The first three terms are the crystal, Hartree and exchange-correlation energy, respectively.
The last three terms coincide with the energy associated with the many-body Hamiltonian
$H_{KS}-H_{DC}+H_U$. the interaction energy $\langle H_{U}\rangle $ may be computed either directly
from the expression of the Hamiltonian $H_U$ (for example, by evaluating the correlations
$\bra n_a n_b\ket$ within the impurity solver, which is easy e.g when using quantum
Monte Carlo). Alternatively, the Migdal formula
$\bra H_U \ket = {\rm Tr}\,(\Sigma G)/2$
can be applied, as actually done in this
work when using the Hubbard-I approximation and described in the next section.
%
%

\subsection{Some practical aspects of total energy calculations}
\label{sec:energy_practical}

An accurate evaluation of the $\vk$-dependent occupancy matrix
 (\ref{occ_mat}) is necessary in order to obtain  the
 correct charge density (\ref{den_mom}) and total energy (\ref{edmft}). If calculations are performed at
finite temperature then this requires a careful summation of
the high-frequency
tails of the Green's function in (\ref{occ_mat}). In addition, the formula  (\ref{edmft}) for the
 total energy contains the Migdal contribution, which is also computed through the corresponding summation
 over the Matsubara frequencies:
 \beq\label{mig_enr}
\langle H_{U}\rangle=\frac{1}{2} {\rm Tr}\left[\hat{G}(\tau=0^-)\hat{\Sigma}(\tau=0^-)\right]=
\frac{T}{2}{\rm Tr} \sum_{n}G(i\omega_n)
\Sigma(i\omega_n)e^{i\omega_n 0^+}
 \eeq
Here both the Green's function and self-energy contain high-frequency tails, which should be properly taken
into account in the evaluation of the sum over $\omega_n$.
In Refs.~[\onlinecite{dei94,dei02,pou05}] evaluation of the Matsubara sums over
the high-frequency tails have been treated in details,
however the approach proposed there is applicable only if the self-energy is calculated by means of an
analytic technique.
Due to the wide spread of numerical DMFT solvers (for example, quantum
Monte Carlo), it is highly desirable
to have a "solver-independent" technique for accurate evaluations of Matsubara sums,
which we describe in this chapter.

We shall isolate the first two terms in the high-frequency expansion of the self-energy
matrix, which is thus decomposed as:
\beq\label{Sig_an}
\hat{\Sigma}(\iomn)= \hat{\Sigma}(i\infty)+\frac{\hat{A}}{\iomn} +
\hat{\Sigma}_{num}(\iomn) \equiv
\hat{\Sigma}_{an}(\iomn) +
\hat{\Sigma}_{num}(\iomn)
\eeq
In this expression, the numerically determined
$\hat{\Sigma}_{num}(\iomn)=\hat{\Sigma}(\iomn)-\hat{\Sigma}_{an}(\iomn)$
will be neglected for Matsubara frequencies larger than a certain cutoff $\omega_{cut}$, while the
high-frequency contribution $\hat{\Sigma}_{an}(\iomn)\equiv \hat{\Sigma}(i\infty)+\hat{A}/\iomn$
will be treated analytically when performing frequency sums. One
may extract the matrices $\hat{\Sigma}(i\infty)$ and  $\hat{A}$ (the latter being actually diagonal in
the present case)
from the real and imaginary parts of the self-energy
at the cutoff frequency $\hat{\Sigma}(i\omega_{cut})$,
 or use a more sophisticated way for fitting them. A similar decomposition can be applied
 to the Green's function:
\begin{eqnarray}
\hat{G}(\vk,\iomn)=&&\hat{G}_{num}(\vk,\iomn)+\hat{G}_{an}(\vk,\iomn)\\
=&&\nonumber \hat{G}_{num}(\vk,\iomn)+\left[(\iomn+\mu)\hat{I}
-\hat{H}^{KS}+\hat{V}^{DC}-\hat{\Sigma}(i\infty) \right]^{-1},
\end{eqnarray}
where
$\hat{G}_{num}(\vk,\iomn)=\hat{G}(\vk,\iomn)-\hat{G}_{an}(\vk,\iomn)$
is again zero for Matsubara frequencies larger than the cutoff
frequency. In the analytical part of the Green's function it is sufficient to keep only the dominant
term in the self-energy $\hat{\Sigma}(i\infty)$.
Then the matrix $\mu+\hat{V}^{DC}-\hat{H}^{KS}-\hat{\Sigma}(i\infty)$ is Hermitian,
and we designate its eigenvectors and
eigenvalues  as $|X_m^\vk\ket$ and $\lambda_m^\vk$, respectively, so that:
\beq\label{G_an}
\hat{G}(\vk,\iomn)=\hat{G}_{num}(\vk,\iomn)+\sum_m \frac{|X_m^\vk\ket \bra
  X_m^\vk|}{\iomn + \lambda_m^\vk}.
\eeq
In order to evaluate the frequency sum in (\ref{occ_mat}) we carry out
the summation of $\hat{G}_{num}$ up to the cutoff frequency, while other frequency sums are
performed analytically.
By inserting (\ref{Sig_an}) and (\ref{G_an}) in the formula for the Migdal energy (\ref{mig_enr})
one obtains the following expressions of practical use:
\begin{eqnarray}\label{e_mig_comp}
&&\langle H_{U}\rangle =\langle H_{U}\rangle ^{(1)}+\langle H_{U}\rangle ^{(2)}+\langle H_{U}\rangle ^{(3)}, \\
&&\nonumber \langle H_{U}\rangle ^{(1)}=\frac{T}{2}\,\sum_{\vk}\,\sum_{\vert\omega_n\vert\langle\omega_{cut}}
\left[\hat{G}_{an}(\vk,\iomn)\hat{\Sigma}_{num}(\iomn)+
\hat{G}_{num}(\vk,\iomn)\hat{\Sigma}(\iomn)\right], \\
&&\nonumber \langle H_{U}\rangle ^{(2)}=\frac{1}{2}\sum_{\vk}\sum_{m}
\frac{\bra X_m^\vk\vert\hat{\Sigma}(i\infty)\vert X_m^\vk\ket}{1+e^{-\beta\lambda_m^\vk}}, \\
&&\nonumber \langle H_{U}\rangle ^{(3)}=
\frac{1}{4}\sum_{\vk}\sum_{m}\frac{\bra X_m^\vk\vert\hat{A}\vert X_m^\vk\ket}
{\lambda_m^\vk}\frac{1-e^{\beta\lambda_m^\vk}}{1+e^{\beta\lambda_m^\vk}},
\end{eqnarray}

The Hubbard-I quantum impurity solver \cite{hubbard_1} employed in the present work is not constrained to the Matsubara
frequencies, and it can in fact be equally well used
for calculating the self-energy at any general complex energy. Hence
at zero temperature one may easily rewrite frequency sums in the expressions for the occupancy matrix (\ref{occ_mat})
and Migdal energy (\ref{edmft}) in
a form, which is suitable for a summation over the poles of
the GF (and the self-energy in the case of Migdal energy)
 on the real axis:
 \beq
N^{\vk}_{LL'}= \frac{1}{2\pi i}\oint G_{LL'}(\vk,z),
\eeq

\beq
\langle H_{U}\rangle=\frac{1}{4\pi i}\oint Tr \left[G(z)\Sigma(z)\right]dz,
\eeq
where integration is performed over the contour in the complex energy plane, which encloses valence-band
energy poles. Therefore, within the Hubbard-I approach and at zero temperature one may completely avoid
the problem of summation of the high-frequency tails. We have applied both the finite temperature Matsubara
temperature summation and contour integration techniques in the fully self-consistent LDA+DMFT calculations
of $\gamma$-Ce and Ce$_2$O$_3$.

\subsection{Choice of interaction vertex, impurity solver, and double-counting}
\label{sec:interaction_solver}

The last term in Eq.(\ref{eq:mb_ham}), $H_U$ gives the many-body interaction terms 
acting in the subset of correlated orbitals. 
They correspond to matrix elements of the Coulomb interaction, and will in general
involve arbitrary 2-particle terms $U_{abcd}c^\dagger_ac^\dagger_bc_dc_c$. There the 4-index matrix 
$U_{abcd}$ is defined for a $f$-shell by four Slatter integrals $F^0$, $F^2$, $F^4$, and $F^6$. In the
quasiatomic (spherical) approximation the Slatter integrals can be expressed through only two parameters
$U$ and $J$ (see Ref.~\onlinecite{anisimov_lda+u_review_1997_jpcm}). In addition to the spherical approximation one often
makes a further simplification and keeps only density-density interactions (we employ this
simplification in the current version of our Hubbard-I impurity solver, though in general it is not necessary). 
We shall limit ourselves here to this case and use:
\beq
H_U =\frac{1}{2}\sum_\vR\sum_{mm'\sigma\sigma'} U_{mm'}^{\sigma\sigma'}\, \hn_{\vR m}^{\sigma}\hn_{\vR m}^{\sigma'}
\label{eq:HU}
\eeq
with the effective two-index matrix derived from a more general 4-index form as follows:
\beq
U_{mm'}^{\spinup\spindown}=U_{mm'mm'}\,\,\,,\,\,\,
U_{mm'}^{\spinup\spinup}=U_{mm'}^{\spindown\spindown}=U_{mm'mm'}-U_{mm'm'm}
\label{eq:U_matrix}
\eeq

Next, we discuss the impurity solver that we use in practice in this article.
Both materials that we shall consider for illustrative purposes
(\ce2o3 and $\gamma$-Ce) have nominally an $f^1$ configuration, and the $f$-electron is
actually localized. The practical solution of the DMFT equation can be
simplified considerably by choosing an approximate `impurity solver' appropriate to
this localized character. The simplest of those is the `Hubbard-I' approximation.
In this approximation, the self-energy is approximated by its `atomic limit', in which
the hybridization function $\Delta_{ab}(z)$ is neglected. One should still correctly
identify however the effective atomic levels entering the Weiss dynamical mean-field
$[{\cal G}_0^{-1}]_{mm'}=z-\epsilon_{mm'}-\Delta_{mm'}(z)$ (with $m,m'\in {\cal C}$).
For this purpose, we can perform a high-frequency expansion of
the self-consistency equation (\ref{eq:scc_esc_dmft},\ref{eq:new_G0}) and request that
$\Delta(z)$ vanishes at high frequency, which leads to:
\beq
\epsilon_{mm'}\,=\,-\mu\,\delta_{mm'}+\sum_\vk H^{KS}_{mm'}(\vk) - V^{dc}_{mm'}
\label{eq:effective_level}
\eeq
In the present case, this is actually a diagonal matrix of effective atomic
levels $\epsilon_{mm'}=\epsilon_m\delta_{mm'}$. These levels must be recalculated iteratively,
as they change upon updating the chemical potential.
Hence, the effective atomic Hamiltonian reads:
\beq
H_{at}^{eff}\,=\,\sum_{\sigma, m\in{\cal C}}\,\epsilon_m\, \hat{n}_{m\s} \,+\, H_U
\label{eq:Hateff}
\eeq
This Hamiltonian can be diagonalized, yielding (many-body) energy levels $E_A$'s
and eigenstates $\vert A\ket$, from which the atomic Green's function can be constructed as:
\begin{equation}
[G_{at}(z)]_{mm'} = \delta_{mm'}\,\frac{1}{{\cal Z}}\sum_{AB}
\frac{|\langle A|d_m^\dagger|B\rangle|^2}{z+E_B-E_A}(e^{-\beta E_A}+e^{-\beta E_B})
\,\,\,,\,\,\,
{\cal Z} = \sum_A e^{-\beta E_A}
\end{equation}
The atomic self-energy is obtained from $\Sigma_{at}=(z-\epsilon_m)\delta_{mm'}-G_{at}^{-1}$.
This leads to the following expression for the Green's function of the full solid,
in the Hubbard-I approximation:
\begin{eqnarray}
[G(\vk,z)^{-1}]_{LL'}\,=&&\,(\mu+\epsilon_L+V^{dc})\delta_{LL'} -
H^{KS}_{LL'}(\vk) + [G_{at}^{-1}(z)]_{LL'}\\
&&= [G_{at}^{-1}(z)]_{LL'} - \widetilde{H}^{KS}_{LL'}(\vk)
\label{eq:Green_Hubbard1}
\end{eqnarray}
in which $\widetilde{H}^{KS}_{LL'}=H^{KS}_{LL'}(\vk)-\bra H^{KS}_{LL'}(\vk)\ket_\vk$
for $L,L'\in{\cal C}$ and $\widetilde{H}^{KS}_{LL'}=H^{KS}_{LL'}(\vk)$ otherwise
(i.e. $\widetilde{H}^{KS}$ is the non-local, inter-atomic part of the KS Hamiltonian).
Within the Hubbard-I approximation, the LDA+DMFT loop takes the following form. Starting from
$H^{KS}$ at a given stage of the iteration, the effective atomic levels are calculated according
to (\ref{eq:effective_level}) and the local atomic Green's function calculated by diagonalizing the
effective atomic Hamiltonian (\ref{eq:Hateff}). The full Green's function is
then formed as (\ref{eq:Green_Hubbard1}) and the momentum distribution matrix
$N^\vk_{LL'}$ calculated. The chemical potential is then updated so that
$\sum_{\vk,L} N^\vk_{LL}$ yields the appropriate total number of electrons, and
the updated charge density is obtained by calculating the moments as described in
Sec.~\ref{sec:density}. The KS equations are then solved to yield a new $H^{KS}$ and
this process is iterated until convergence of both $\rho(\vr)$ and of the effective
atomic levels $\epsilon_m$.

Finally, we discuss the ``double-counting'' correction $H_{DC}$. This correction
must be introduced, since the contribution of interactions between the
correlated orbitals to the total energy is already partially included in
the exchange-correlation potential derived from $E_{xc}$.
Unfortunately, it is not possible to derive this term explicitly, since the energy
within DFT is a functional of the total electron density, which combines all orbitals
in a non-linear manner. In practice, the most commonly used form of the double-counting term
is (for other choices, see e.g~[\onlinecite{lichtenstein_magnetism_dmft_2001_prl}]):
\begin{eqnarray}\nonumber
H_{DC} = \sum_{\vR\sigma ab} V^{DC}_{ab\sigma} c^\dagger_{a\sigma}c_{b\sigma}\\
V^{DC}_{ab\sigma} = \delta_{ab}\left[U (N_f-\frac{1}{2})-J(N_f^\sigma-\frac{1}{2})\right],
\label{eq:dc_potential}
\end{eqnarray}
where $N_f=N_f^\uparrow+N_f^\downarrow$ is the total occupancy in the correlated shell
${\cal C}$ (i.e the $f$-shell in practice in this article).
One issue arises here, which is which value of $N_f$ must actually be used in
(\ref{eq:dc_potential}). When solving the DMFT equations with a numerically exact solver,
it would seem (from the previous functional-based derivation) that $N_f$ should be
the occupancy of the $f$-shell obtained at self-consistency. However, we have found this
to be inappropriate when using the Hubbard-I approximation as a solver and leading to
too small equilibrium volumes. Instead, we note that the Hubbard-I solver treats
an effective isolated atom, which has this a frozen occupancy taking integer
values at $T=0$. Hence, a natural choice within Hubbard-I is to choose this
frozen integer occupancy in the double-counting correction. For the $f^1$ materials
in the paramagnetic phase treated in this article
($N_f^\uparrow=N_f^\downarrow=1/2$), so that the Hund's coupling therefore
drops out of $H_{DC}$ which reads simply:
\begin{eqnarray}\nonumber
V^{DC}_{ab\sigma} = \delta_{ab}\frac{U}{2},
\end{eqnarray}
This also implies that the $U$-dependent contribution of the double counting
correction in the total energy actually vanishes for $f^1$ compounds within the Hubbard-I
approximation, and for this choice of double-counting potential.

\section{Fully self-consistent LDA+DMFT calculations of Ce$_2$O$_3$ and $\gamma$-Cerium}
\label{sec:results}

We have applied the LMTO-based fully self-consistent LDA+DMFT technique
in order to calculate the density of states and thermodynamical properties of the
cerium sesquioxide Ce$_2$O$_3$ and pure Ce in its $\gamma$-phase. The main aim of these
calculations is to validate
our implementation of the fully self-consistent LDA+DMFT technique, as well as to study
the impact of the charge self-consistency
on spectral and thermodynamic properties. Hence, it is desirable to avoid additional
complications due to the use of sophisticated and computationally expensive quantum impurity
solvers. While the Ce 4$f$
band cannot be properly modeled by conventional LDA techniques, due
to its essentially atomic-like (localized) character, the electronic correlations on the 4$f$ shell
can be treated by means of the simplest "strong-coupling" Hubbard-I (HI)~\cite{hubbard_1}
approximate impurity solver, described in the previous section.
Therefore for our purposes
these two strongly correlated rare-earth compounds, with nominally one localized $f$-electron,
appear to be a suitable choice. Both pure cerium and the oxide Ce$_2$O$_3$ are well studied theoretically
and experimentally, because they raise questions of fundamental interest
(e.g. concerning the $\alpha$-$\gamma$ transition in Ce \cite{kosk78,mcmahan_collapse_review_pub}) as well as
for technological reasons (for example, Ce-oxides are used for
oxygen storage in solid-oxide fuel cells \cite{skr01}).

\subsection{Ce$_2$O$_3$}
\label{sec:ce2o3}

The electronic properties of Ce$_2$O$_3$ are largely determined by the
localized Ce $4f$ orbitals. This compound is an insulator with a gap of
about 2.5 eV. At low temperatures, Ce$_2$O$_3$ orders
antiferromagnetically, with a N{\'e}el temperature (9 K) which is
several orders of magnitude smaller than the gap. It is thus clear that the antiferromagnetic
order
is not the driving force behind the insulating character of this material. Rather,
strong correlations open up a gap among the $4f$ states, and this compound can
therefore be viewed as an f-electron based Mott insulator.
As for all such materials, it is a challenge
for conventional electronic structure calculations to describe the opening of the
correlation-induced gap. The Kohn-Sham spectrum of DFT-LDA is metallic, with $4f$
states at the Fermi level. An insulating state of Ce$_2$O$_3$ has been obtained within the self-interaction correction approach \cite{petit07}, albeit with a strongly overestimated value for the band gap.
In the antiferromagnetic phase, the LDA+U method can be
employed \cite{fabris05,singh06,ander07,losch07}, as well as the hybrid functional approach \cite{silva07}.
In this article, we focus on the paramagnetic phase, at low temperature just above
the N\'eel ordering temperature. Dynamical mean-field theory and the Hubbard-I
approximation used here allows us to describe the opening of the Mott gap and the
existence of local moments in the paramagnetic phase, due to the quasi-localized
$4f$ electrons.

\subsubsection{Details of the calculational setup}

We have calculated the density of states and spectral density of
 Ce$_2$O$_3$ oxide in the hexagonal lattice structure (space group $P\bar{3}m1$)
at experimental volume ($a=$3.890 \AA, $c/a$=1.557 \cite{wyc67}). Empty spheres have been introduced in order
to make the structure more close packed. In the computation of the spectra we have included the
$6s$, $5d$ and $4f$ orbitals of Ce,
the $2p$ orbitals of oxygen and the $1s$ orbitals on the empty spheres in the basis set as active LMTOs. In addition,
the $6p$ orbital of Ce, the $3s$ and $3d$ oxygen orbitals, and the $2p$ orbitals on the empty spheres have been
downfolded. In the total energy calculations it is necessary to
 include the Ce semicore 5$p$ orbital
\cite{skr01} as well. The LMTO code employed by us is not able to treat semicore states in a separate panel, hence
 in the total energy calculations we have excluded the Ce 6$p$
 orbitals from the basis. It is important to include
the 6$p$
orbitals in order to obtain accurate spectra; however, as will be shown in the next chapter on $\gamma$-Ce,
for total energy calculations they are less significant.

In the self-consistent LDA+DMFT calculations we have employed 62 $\vk-$points
in order to carry out the integration over the irreducible Brillouin zone of the
hexagonal lattice, the energy integration has been carried out on a semicircular contour of depth 2 Ry
comprising 40 energy points. In order to
obtain the local and $\vk$-resolved spectral functions,
the self-energy and Green's function
were directly computed on the real axis, with
394 $\vk-$points used for the integration over the irreducible Brillouin zone.
%
%
%


In order to estimate the value of the screened on-site Coulomb interaction we have performed constrained
LSDA calculations \cite{anisimov_calcU_1991_prb} both, by fixing the $f-$electron
occupancy and by adding a constrained
potential acting on the $f$-electrons on one Ce atom. Depending on whether the constrained
charge or potential approach was used we obtained quite different values of the parameter $U$, varying from 5.5 to 8 eV.
The previous theoretical estimates of Refs.~[\onlinecite{herb78,mcmahan_collapse_review_pub}] of the
$U$ value in pure Ce metal put it in the range between 5 and 6 eV, while in the LDA+U calculations of
Refs.~[\onlinecite{ander07,losch07}] the best agreement between the calculated
equilibrium volume and experiment was achieved for a value of the effective $U_{\rm{eff}}= U-J$ of about 6~eV.
Thus, in our calculations we have fixed the value of $U-J$ to 6 eV, while the value of $J$=0.46~eV,
which depends less crucially than $U$ on the proper treatment of screening in a material, has been taken
from the constrained LDA calculations.

We have employed three calculational schemes, for comparison purposes: the conventional LDA, the (non self-consistent)
LDA+DMFT with fixed LDA charge density (DMFT-nonSC), and the LDA+DMFT technique with full
self-consistency over the charge density (DMFT-SC). In the case of DMFT-nonSC,
we carried out DMFT iterations until convergence in the self-energy and chemical potential
is reached. In the case of DMFT-SC, additional convergence criteria with respect to the total
energy and charge density were employed as well.

\subsubsection{Spectral functions and correlated `bandstructure'}

\begin{widetext}
\begin{center}
\begin{figure}[t]
\includegraphics[width=9 cm]{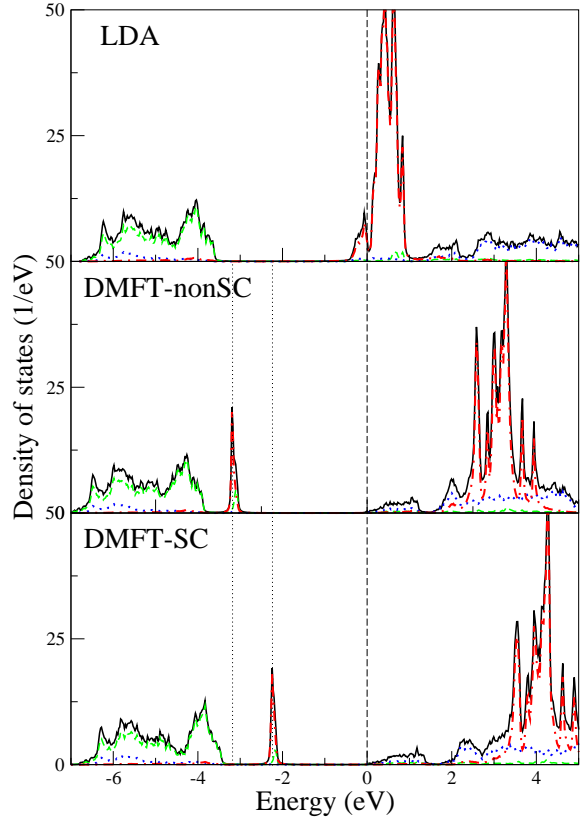}
\caption{(Color online). The LMTO Ce$_2$O$_3$ DOS calculated within DFT-LDA, the LDA+DMFT with the fixed LDA charge density (DMFT-nonSC),
and
the LDA+DMFT fully self-consistent over the charge density (DMFT-SC) schemes.
The black (solid), green (dashed), blue (dotted) and red (dash-dotted) curves are
the total, O, Ce$-d$ and Ce$-f$ DOS, respectively. The vertical
dashed line indicates the position of the chemical potential, while the vertical dotted lines indicate positions of the
lower Hubbard bands in DMFT-nonSC and DMFT-SC, respectively.}
\label{Ce2O3_dos}
\end{figure}
\end{center}
\end{widetext}

The orbitally-resolved local density of states (DOS) - or spectral functions-, defined as:
\beq
A_{L}(\omega)\,\equiv\,-\frac{1}{\pi}\,\rm{Im}\,
\sum_\vk\,G_{LL}(\vk,\omega+i0^+)
\label{eq:def_dos}
\eeq
are displayed in Fig.~\ref{Ce2O3_dos} for all three methods (LDA, DMFT-nonSC, and DMFT-SC).
First, one may notice that conventional LDA calculations place the Ce $4f$-band in the vicinity of
the Fermi level, therefore predicting Ce$_2$O$_3$ to be a metal.
Both the DMFT-nonSC and DMFT-SC approaches correctly predict Ce$_2$O$_3$ to be a Mott insulator, where the $f$-band
is split due to the local Coulomb interaction into occupied lower Hubbard bands, and empty upper Hubbard bands.
One may notice a significant shift of the positions of the Hubbard bands in the DMFT-SC DOS with respect
to the DMFT-nonSC picture. The value of the band gap in Ce$_2$O$_3$ is equal to $3.10$~eV
and $2.13$~eV within the DMFT-nonSC and DMFT-SC approaches, respectively.
The latter value is in good agreement with the experimental measurements
of the optical gap in Ce$_2$O$_3$ ($2.4$~eV, Ref.~[\onlinecite{Golub95}]), while
the fixed charge (non-SC) calculations lead to a strong overestimation of the gap.
The total occupancy of the $f$-shell is rather weakly affected
by the local Coulomb interaction. The occupancy of the $f$ shell in Ce$_2$O$_3$ is equal to
$1.145$, $1.174$, and $1.167$
according to the LDA, DMFT-nonSC and DMFT-SC calculations, respectively.

\begin{table}
\begin{tabular}{l|ccccc|}
\hline
 &\hm $\mu$ & $C^f_{KS}$  &\hm $\langle H^f \rangle$ &\hm $V_{DC}$ &\hm $\langle\epsilon_f\rangle$ \hm\\
\hline
\hm DMFT-nonSC \hm & \hm 1.19 &\hm  1.52 &\hm 1.18 &\hm -3.00 &\hm  -3.01 \hm \\
\hm DMFT-SC \hm & \hm 1.15 &\hm  2.76 &\hm 2.18 &\hm -3.00 &\hm  -1.97 \hm \\
\hline
\end{tabular}
\caption{The chemical potential, parameter $C$ of the KS $f$ band, double counting and $f$ level positions in the
DMFT-nonSC and DMFT-SC approaches (in eV).}
\label{table_compar}
\end{table}

In order to better understand the observed difference between the position of the
lower Hubbard band in the non-SC and SC methods, we use the expression (\ref{eq:effective_level})
of the effective atomic levels within the Hubbard-I approximation, established in the previous
section. This yields directly the position of the lower Hubbard band (and of the upper Hubbard band,
which is shifted upwards by an energy $U$) since, for both materials, the $f$ shell of the isolated atom
contains just one electron.
The effective level position for the $m$-th orbital
component of the $4f$ shell reads:
\begin{equation}\label{e_l}
\epsilon_m=-\mu+\sum_{\vk}\hat{H}_{mm}^{KS}({\bf k})-V_{DC},
\end{equation}
The double counting term used in the present work reads:
$V_{DC}=U(N_f-1/2)-J(N_f-1)/2$ and since we set $N_f=1$ in this expression (as explained in the previous
section), we see that the double counting correction term is the same for both the
DMFT-nonSC and DMFT-SC approaches. Thus, the difference in the value of the band gap
cannot be blamed on the double counting.
Therefore, in the Hubbard-I approximation, and for a given local
Coulomb interaction, the factor which determines the positions of the lower Hubbard band
is the momentum average of $H_{KS}$, i.e essentially the centre of gravity of the KS f-band
(the second term in the right-hand side of expression
(\ref{e_l})).
\begin{widetext}
\begin{center}
\begin{figure}[t]
\includegraphics[width=10 cm]{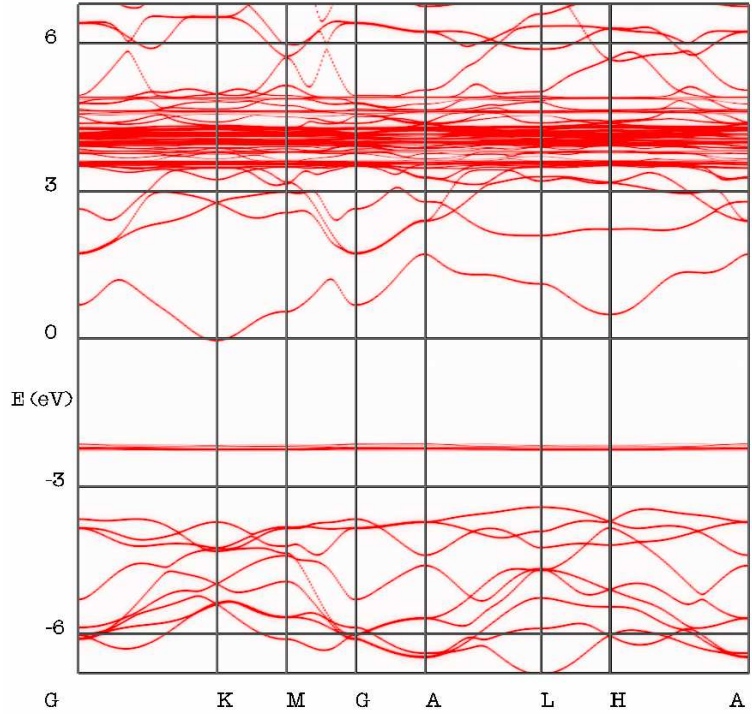}
\caption{(Color online). The LMTO fully self-consistent LDA+DMFT {\bf k}-resolved spectral function of Ce$_2$O$_3$.}
\label{Ce2O3_bands}
\end{figure}
\end{center}
\end{widetext}
In Table~\ref{table_compar} we list the chemical potentials, the LMTO parameter
$C^f$ of the Ce $f$ band (which can be interpreted as the center of a "pure" (unhybridized) KS band),
the Hamiltonian integrated over
the BZ, the value of the double counting correction and the
level position. The brackets $\langle ... \rangle$ in
Table~\ref{table_compar} designate an average over
the magnetic quantum number $m$ within the f-shell.
It is important to realize, that while in the case of DMFT-nonSC, the
parameters $C^f_{KS}$ and $H^f$ are just obtained within the conventional LDA,
in DMFT-SC it is obtained from the solution of the KS equations with a potential
calculated from the self-consistent LDA+DMFT charge density, which differs from the
LDA one.
%
One may notice that the KS band structure computed from the DMFT charge density is substantially
different from the LDA one, as the centerweight of the $f$ band is shifted by
$1.24$~eV upwards.
This change in the KS band structure is reflected in the values of the Hamiltonian $\langle H^f \rangle$, and,
finally, in the positions of the DMFT Hubbard bands, which are determined by the level position
$\epsilon_m$. Hence one observes a clear
impact of the charge-density self-consistency on the LDA+DMFT
electronic structure: changes in the charge density
due to DMFT lead to a modification of the KS band structure, which in turn affects the final LDA+DMFT electronic
structure. The result is a reduction of the band gap by almost $1$~eV.
A direct connection between the KS and DMFT electronic structures is especially evident in the present
case due to the simplicity of the HI technique, where the position of the LHB is linked to the KS Hamiltonian through
the simple relation (\ref{e_l}).

Since the Hubbard-I approximation employed here does not describe
lifetime effects of the correlated states, the momentum-resolved
Green's function still has well-defined poles (in other words, excitations
corresponding to the lower and upper Hubbard bands have infinite lifetime).
The dispersion of the corresponding upper and lower Hubbard bands of \ce2o3 as a
function of momentum is displayed in Fig.~\ref{Ce2O3_bands}, as obtained with
fully self-consistent LDA+DMFT.
The LHB shows virtually
no dispersion and forms a narrow, atomic-like level in the gap between the oxygen $p$ and Ce $d$ band. The UHB
is essentially a set of atomic-like "multiplet" states, while one
may still notice some evidence of weak hybridization between the $d$
band and the UHB \footnote{One may notice that in Fig.~\ref{Ce2O3_bands} the chemical potential is pinned little bit above the bottom of the conduction band. This is actually a drawback of the Hubbard-I approximation, in which the insulating
plateau on the $n(\mu)$ curve is not exactly at a commensurate filling (due to the non-conserving nature of the 
Hubbard-I approximation). This is a very small effect, however (of order of 0.01\% of the total electronic charge),
therefore producing a negligible change in the total charge density.} .

\subsubsection{Total energy calculations}

The total energy calculations have been carried out by keeping the experimental $c/a$ ratio and
using the same setup as in the calculations of the spectra, apart
from modifications in the basis, where the 5$p$ Ce semicore states were included instead of the 6$p$ orbitals.
Without the 5$p$ semicore states there is no minimum of the total energy vs. volume curve in the range of lattice
parameters between 3.60 and 4.2 \AA. As in the calculations
of spectra, we have employed both the fully self-consistent and fixed LDA charge density implementations of
the LDA+DMFT method. In order to compare our results obtained within the LMTO-ASA with the
recent full-potential calculations of Andersson {\it et al.}~\cite{ander07}, we have also carried out LDA and LDA+U
calculations of Ce$_2$O$_3$. The LDA+U calculations have been performed using the same setup and $U$ and $J$
parameters than for LDA+DMFT, but using a Hartree-Fock approximation to the self-energy in
the self-consistent calculations. The HF self-energy is obtained for an $f$ occupancy,
which can be different from the atomic one, therefore, in
contrast to the HI case, in the LDA+U calculations
we have used the actual (self-consistent)
$f$-band occupancy in the expression for the double-counting correction.
\begin{figure}[t]
\includegraphics[width=7 cm, angle=270]{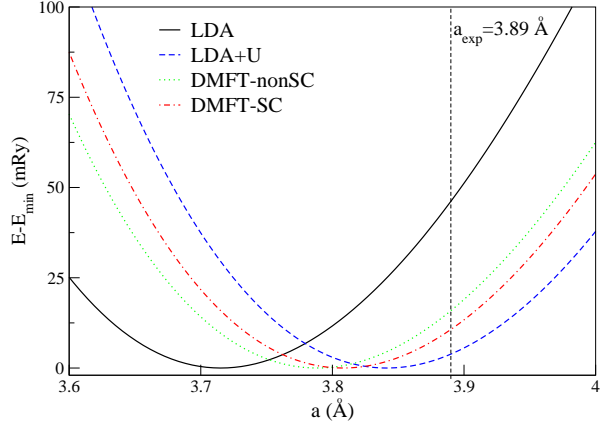}
\caption{(Color online). The total energy vs. lattice parameter dependence calculated within
LDA (solid line), LDA+U (dashed line), DMFT-nonSC (dotted line) and DMFT-SC (dash-dotted line) approaches.The vertical
dashed line indicates the experimental lattice parameter of Ce$_2$O$_3$}
\label{Ce2O3_ener}
\end{figure}

Our results for the energy vs. lattice parameter curves are displayed in
Fig~\ref{Ce2O3_ener}. One may see that the localization of the Ce 4$f$ electrons
due to the strong local Coulomb correlations leads to a substantial increase in the lattice parameter.
This is expected on a physical basis, since localization leads to a
decreasing participation of these electrons to the cohesion of the solid.
Our results for the equilibrium lattice parameter within the DMFT-nonSC,
DMFT-SC and LDA+U methods are rather close to each other
(3.79, 3.81 and 3.84 \AA), and
in reasonable agreement with the value 3.86 \AA, obtained in
Ref.~[\onlinecite{ander07}] for a similar value of $U$, as well
as with experiment (3.89 \AA). The LDA lattice constant is much smaller (3.72~\AA).
The difference between the DMFT-nonSC and  DMFT-SC for the equilibrium volume appears,
somewhat surprisingly, to be rather small. The difference
between the LDA+U and LDA+DMFT equilibrium volumes
(i.e between Hartree-Fock and Hubbard-I approximations for the self-energy) stems
possibly from the non-conserving nature of the HI approximation.

\subsection{$\gamma$-Cerium}
\label{sec:cerium}

\begin{widetext}
\begin{center}
\begin{figure}[h]
\begin{tabular}{cc}
{\resizebox{6.0cm}{!}
{\rotatebox{0}{\includegraphics{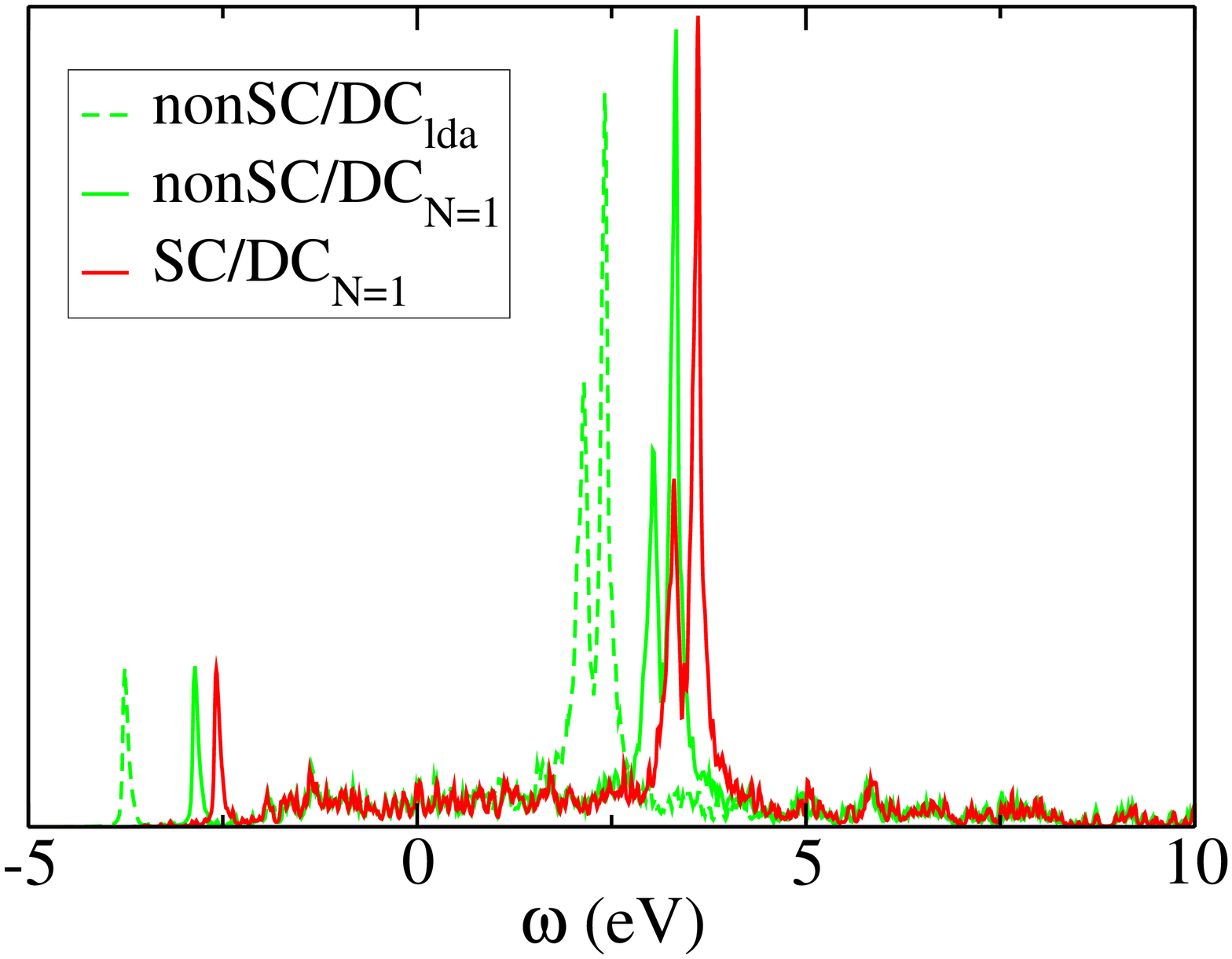}}}} &
{\resizebox{6.0cm}{!}
{\rotatebox{0}{\includegraphics{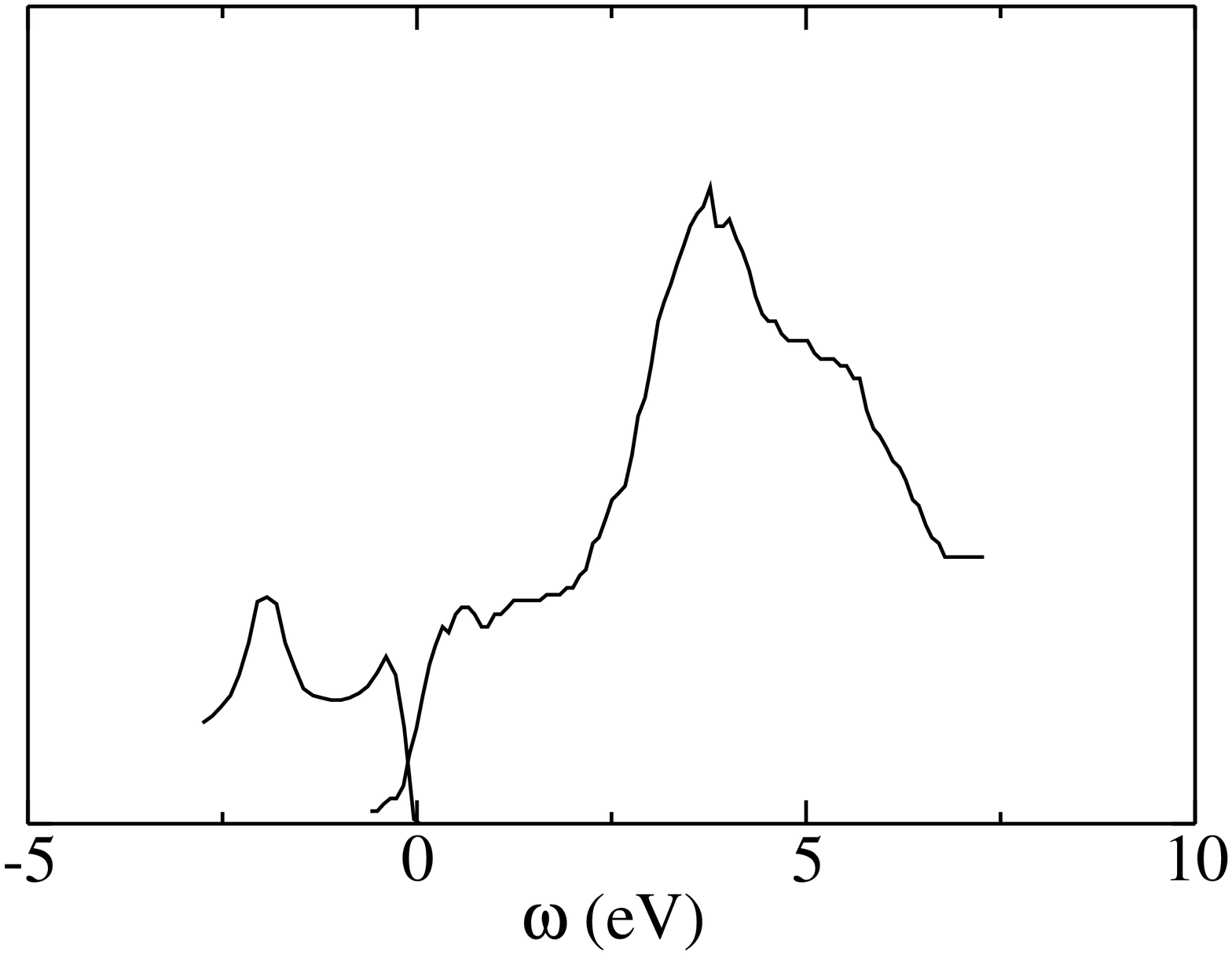}}}} \\
\end{tabular}
\caption{(Color online). Spectral function for $\gamma$-Cerium within LDA+DMFT (Hubbard I) at T=0K
with the self-consistent (SC) and non self-consistent (nonSC) schemes. In the first case,
two ways of treating the double counting corrections are used: LDA number of electron 
and DMFT one (N=1 for the Hubbard I solver). Experimental spectrum (PES and BIS) of $\gamma$ 
Cerium\cite{1983PhRvB..28.7354W,1982PhRvB..26.7056W} are also represented.}
\label{Ce_spectres}
\end{figure}
\end{center}
\end{widetext}


As a second example,
we have applied our DMFT-SC code to fcc-Cerium.
This system has been thoroughly studied in the past
in order to understand the $\alpha$-$\gamma$ transition \cite{kosk78},
an isostructural volume collapse transition taking place as
a function of temperature. Above the transition temperature T$_c$,
in the large volume ($\gamma$-) phase, the f-electrons are
localized, while the smaller volume in the $\alpha$-phase below
T$_c$ leads to a more delocalized behavior of the f-electrons.
Several model studies have been carried
out to describe this transition
\cite{johansson_1974_mott_philmag,lavagna_kvc_physlett_1982,liu_allen_cerium_spectro}.
More recently,
it has also been studied in the framework of ab-initio calculations
using LDA+DMFT~\cite{1999PhRvB..59.3450L, 2001PhRvL..87A6403Z, held_cerium_2001_prl,
held_cerium_2003_prb, MacMahan2005, haule_cerium_prl_2005,2005JPSJ...74.2517S, ama06}.
However, these studies use the fixed charge density (non-SC) implementation of LDA+DMFT.
Nevertheless,
they were successful in describing the main aspects of the transition (spectra,
and Kondo stabilization energy).
The goal of our study is to evaluate
the importance of self-consistency effects in LDA+DMFT calculations for Cerium.
Since the most recent calculations  \cite{held_cerium_2003_prb,held_cerium_2001_prl,ama06} were done with the precise but time consuming Quantum Monte Carlo,
it was not possible to test it.
Another important goal is to study if a simple
Hubbard I implementation of LDA+DMFT is able to describe
correctly the $\gamma$ phase of Cerium, in which
electrons are more localized than in $\alpha$ Cerium.

Computations were done at the experimental volume of the $\gamma$ phase for the spectra
(34.8${\rm \AA}^3$). 5p semicore states were taken into account thanks to an
implementation of the LDA+DMFT self-consistency over density scheme with the
multiple LMTO code \cite{1994PhRvB..49.7219A}. Calculations with 5s states included do not show any change
in the results. Valence states contain 5p 6s 6p 5d and 4f states. Calculations were done
with 145 $\vk$-points in the irreducible Brillouin zone.

\begin{table}[b]
\begin{tabular}{l|ccccc|}
\hline
 &\hm $n_f$ &\hm $\mu$ & \hm $\langle H^f \rangle$ &\hm $V_{DC}$ &\hm $\langle\epsilon_f\rangle$ \hm\\\hline
\hm DMFT-nonSC/DC$_{\rm LDA}$ \hm &1.06 &\hm 0.23 &\hm  0.45  &\hm 3.94  &\hm  -3.72 \hm \\
\hm DMFT-nonSC/DC$_{\rm N=1}$ \hm &1.04 &\hm 0.26 &\hm  0.45 &\hm 2.99 &\hm  -2.80 \hm \\
\hm DMFT-SC/DC$_{\rm N=1}$ \hm &1.04 &\hm 0.10 &\hm  0.57 &\hm 2.99 &\hm  -2.52 \hm \\
\hline
\end{tabular}
\caption{The number of electrons in DMFT, the chemical potential, the double counting
 contribution to the potential  and  the $f$ level position in the
 DMFT-nonSC and DMFT-SC approaches for Cerium. Energies are in eV.
 The number of f-electrons in LDA is 1.16.}
\label{table_dos}
\end{table}

Figures \ref{Ce_spectres} and  \ref{Ce_energy} shows the spectral function and the
total energy versus volume curves
for fcc-Cerium for both the non-SC and
the SC calculations. In accordance to calculations on Ce$_2$O$_3$, we use the `atomic' number
of electron computed  in the Hubbard I solver to evaluate the double counting correction.
It is more justified because of error compensation and, as we show below, the results for the
lattice parameter at T=0 are consistent with
LDA+U self-consistent calculations using the self-consistent number of electrons in the solid.
For testing purposes, a non-SC calculations has also been done with the LDA number of
electron in the double-counting correction.

Concerning the spectral function, we first see  that all the calculations
give the same qualitative physical pictures. The distance in energy between Hubbard bands
is the same in all the calculations. The difference lies in the exact position of the Hubbard bands.
The SC calculation gives the best agreement with experiment, the non-SC is shifted by 0.2 eV.
The non-SC calculation with LDA double-counting is shifted by 1 eV with respect to the SC calculation.
Table \ref{table_dos} gives the data necessary to understand these shifts, as explained before.
We see that the effect of self-consistency itself is small in this case.
The main error in the non-SC/DC$_{\rm LDA}$ comes from the double counting term, as expected.
Note that non-SC and SC calculation lead to nearly the same number of electron.
This can be partly attributed to the correlated character of the $\gamma$ phase of Cerium:
electrons are localized, so the number of electron is close to 1 (at least compared to
an LDA calculation).
Experimental spectra\cite{1983PhRvB..28.7354W,1982PhRvB..26.7056W} of the gamma phase are also
represented on figure \ref{Ce_spectres}. While the calculated position of the Hubbard bands is
in good agreement with experiment, the width of the upper Hubbard band is not correctly
described in our calculations. This is not surprising and is to be blamed
on the use of a simplified Hubbard-I solver, which does not include
lifetime effects.
This means that Hubbard-I is only partly adequate to describe
the high temperature $\gamma$-phase.
\begin{widetext}
\begin{center}
\begin{figure}[t]
\includegraphics[width=9 cm]{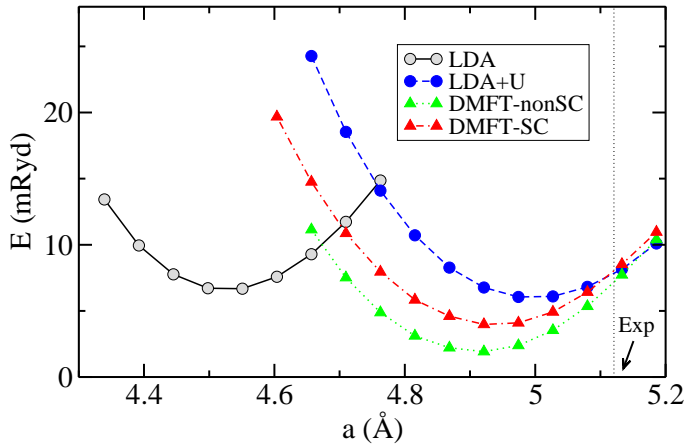}
\caption{(Color online). Total energy versus lattice parameter curves in LDA+DMFT (Hubbard I) for Cerium (shifts in energy between different curves are arbitrary).}
\label{Ce_energy}
\end{figure}
\end{center}
\end{widetext}

Figure \ref{Ce_energy} gives the total energy versus volume computed at T=0K.
The main conclusion is that non-SC and SC DMFT calculations give the same minimum for
the internal energy versus volume curves (see also table \ref{tab:compa}). This value is not far from
the value obtained in the LDA+U calculations in ASA (which is itself
in good agreement with LDA+U calculations in PAW \cite{amadon07}).
Note however that LDA+U can only account for correlation effects by introducing
a spurious magnetic order, while DMFT is able to describe local
moment formation and correlation effects in the paramagnetic phase.
Effect of self-consistency appears to be fairly weak in the case of $\gamma$-cerium. 
One should also keep in mind that the ASA approximation
does not describe correctly the bulk modulus in LDA. So the results
concerning the energy versus volume curves should
be taken only as trends.
Moreover, direct comparison with experiment is difficult,
because we have not computed the full
free energy but only the internal energy~\cite{ama06}.

%


\begin{table}[h]
\begin{tabular}{lc}
\hline
 Exp\cite{Jeong04,Olsen85}&  5.16 \\
 LDA+U/PAW\cite{amadon07}                &  5.05 \\
 LDA+U/ASA                &  5.00 \\
 DMFT SC/ASA              &  4.93 \\
 DMFT non-SC/ASA          &  4.91 \\
LDA/PAW \cite{amadon07}        &  4.52 \\
LDA/ASA         &  4.52 \\
\hline
\end{tabular}
\caption{Lattice parameter (in \AA) of $\gamma$ Cerium according to
experimental data, PAW calculations and our calculations.}
\label{tab:compa}
\end{table}

The calculations where 6p states are not taken into account
leads to an underestimation of the lattice parameter
(4.80 \AA\  instead of 4.93 \AA).
This is in accordance to the underestimation
of the lattice parameter in LDA
when 6p states are neglected (4.44 \AA\ instead of 4.52 \AA )

Our conclusion regarding $\gamma$-cerium is that spectra and energy are reasonably
described with a fixed (LDA, non-SC) charge density approximation. However,
it is mandatory to use a self-consistent implementation to obtain
more accurate results.


\section{Conclusion}
\label{sec:conclusion}

In conclusion, we have devised a simple and efficient implementation of the fully self-consistent
LDA+DMFT method in the LMTO basis set. The charge density is calculated from moments involving
the LDA+DMFT momentum-distribution matrix and the KS hamiltonian. We have also
obtained accurate formulas for
computing the total energy by handling high-frequency
tails of the Green's function and self-energy in an appropriate manner.

We have computed  the local and $\vk$-resolved spectral functions of
cerium sesquioxide Ce$_2$O$_3$ by means of the fully self-consistent
 LDA+DMFT technique in conjunction with the Hubbard-I
approximation for the DMFT self-energy.
We have shown that the charge-density self-consistency affects the spectral properties of
 Ce$_2$O$_3$ substantially, causing a shift of the lower Hubbard band
 by approximately $1$~eV and a corresponding decrease
in the value of the band gap in comparison with DMFT calculations with
fixed LDA charge density;
This effect considerably improves the agreement with
experiment in comparison to a non-self consistent calculation using
the LDA charge density. We have identified the main cause of these modifications, which is
due to a significant change in the effective KS band structure, computed with
LDA+DMFT charge density, as compared to the
LDA band structure.

Finally, we have obtained the total energy and equilibrium volume of $\gamma$-Ce
within the fixed (LDA) charge and fully self-consistent LDA+DMFT schemes. The effects of the
self-consistency over the charge density are less important for
$\gamma$-Ce than for the oxide: self-consistency induces in this case
a change of  $1$\% on the ground-state volume, and affects the spectral
function by shifts of a fraction of an electron-Volt only.

\section*{ACKNOWLEDGMENTS}
                                                                                                                                                             
We are grateful to F. Aryasetiawan and R. Windiks for useful discussions, and to A. Poteryaev for providing us with a Hubbard-I quantum impurity solver computing code. 
Financial support from CNRS, \'{E}cole Polytechnique and the E. U. "Psi-k f-electron"
Network under contract HPRN-CT-2002-00295 is acknowledged. This work was supported by "Materials Design", Le Mans, and 
by a supercomputing grant (No. 071393) at IDRIS, Orsay.

\bibliographystyle{apsrev}

\end{document}